\documentclass[aps,prl,twocolumn,superscriptaddress]{revtex4-2}

\usepackage{graphicx}
\usepackage{dcolumn}	
\usepackage{bm}			
\usepackage{here}
\usepackage{color}
\usepackage{xcolor}
\usepackage[normalem]{ulem}
\usepackage{amsfonts}

\begin{document}


\title{Strain-Tuned Nodal Superconductivity in the Charge-Ordered Kagome Metal CsV$_3$Sb$_5$}


\author{Yusuke Takeuchi}
\affiliation{Department of Physics, Okayama University, Okayama 700-8530, Japan}

\author{Akito Kobayashi}
\affiliation{Department of Physics, Okayama University, Okayama 700-8530, Japan}

\author{Saki Uchida}
\affiliation{Department of Physics, Okayama University, Okayama 700-8530, Japan}

\author{Takumi Nagao}
\affiliation{Department of Physics, Okayama University, Okayama 700-8530, Japan}

\author{Seigo Ogawa}
\affiliation{Department of Physics, Okayama University, Okayama 700-8530, Japan}

\author{Rui Zhou}
\affiliation{Institute of Physics, Chinese Academy of Sciences, and Beijing National Laboratory for Condensed Matter Physics, Beijing, 100190, China}

\author{Shinji Kawasaki}
\email[]{kawasaki@science.okayama-u.ac.jp}
\affiliation{Department of Physics, Okayama University, Okayama 700-8530, Japan}

\author{Fei Song}
\affiliation{Fujian Provincial Collaborative Innovation Center for Advanced High-Field Superconducting Materials and Engineering, College of Physics and Energy, Fujian Normal University, Fuzhou 350117, China}

\author{Hao Ni}
\affiliation{Fujian Provincial Collaborative Innovation Center for Advanced High-Field Superconducting Materials and Engineering, College of Physics and Energy, Fujian Normal University, Fuzhou 350117, China}

\author{Yong Zhao}
\affiliation{Fujian Provincial Collaborative Innovation Center for Advanced High-Field Superconducting Materials and Engineering, College of Physics and Energy, Fujian Normal University, Fuzhou 350117, China}
\affiliation{Guang'an Institute of Technology, Guang'an, Sichuan 638000, China}

\author{Guo-qing Zheng}
\email[]{zheng@psun.phys.okayama-u.ac.jp}
\affiliation{Department of Physics, Okayama University, Okayama 700-8530, Japan}


\date{\today}

\begin{abstract}

The nature of the superconducting pairing symmetry in the kagome metal CsV$_3$Sb$_5$ and its relationship with the charge density wave (CDW) order are central unresolved issues. Here, we investigate the evolution of superconductivity in CsV$_3$Sb$_5$ under in-situ uniaxial pressure using $^{121}$Sb nuclear quadrupole resonance (NQR). We find that tensile strain significantly enhances the superconducting transition temperature, $T_{\rm c}$, while the CDW remains unchanged, demonstrating that superconductivity can be tuned independently of the bulk charge order. At a tensile strain of $\varepsilon$ = +0.90\%, the nuclear spin-lattice relaxation rate reveals a remarkable double transition: an upper transition at  $T_{\rm c1}$ = 3.6 K to a nodal gap state, and a lower one at $T_{\rm c2}$ = 3.0 K characterized by a nodeless gap. These results evidence degenerate superconducting states with different gap symmetry in the kagome metal at ambient pressure which split under strain. Our work demonstrates a high tunability of superconductivity by uniaxial pressure.

\end{abstract}


\maketitle

\clearpage

Unconventional superconductivity in strongly correlated electron systems often appears near a quantum critical point of an electronic or spin order \cite{Dagotto}. The quintessential examples are the $d$-wave high-temperature superconductivity in cuprates \cite{Keimer} and the complex pairing in pnictides \cite{FernandesFe}, both emerging from the correlated 3$d$ electrons on square lattices. In these materials, disentangling the relationship between superconductivity and intertwined background states, such as antiferromagnetism and the charge density wave (CDW) and/or nematic order, has been a central theme of research \cite{KitahashiNatCom,ZhouNatCom}. Beyond the square lattice, systems with special crystal geometries are now drawing significant attention as platforms for exotic electronic phenomena. For example, materials with trigonal, honeycomb or kagome lattices have been considered as spin liquid candidates due to geometric frustration \cite{QSL,FengSL,Lee_NatPhys}.

The kagome metal CsV$_3$Sb$_5$ represents a recent new frontier \cite{Ortiz,OrtizReview}. It features a perfect two-dimensional kagome network of vanadium (V) atoms, a unique lattice geometry that gives rise to a characteristic electronic structure with flat bands, van Hove singularities (vHs), and Dirac cones \cite{OrtizReview}. This system exhibits a cascade of symmetry-breaking phases, beginning with a structural phase transition into an unconventional CDW state at $T_{\rm CDW}$ $\sim$ 94 K, which is followed by a superconducting ground state at $T_{\rm c}$ $\sim$ 2.5 K \cite{Ortiz,OrtizReview}. The CDW state itself is highly exotic, with experimental reports of time-reversal and rotational symmetry breaking \cite{ZWangSTM,OrtizPRX,ZShan_muSR,LuoNPJ}, suggesting a complex order parameter beyond a simple periodic lattice distortion \cite{TazaiPNAS}.

A central, unresolved question is the symmetry of the superconducting gap and its relationship with the competing CDW order. The pressure-temperature phase diagram, where $T_{\rm c}$ exhibits a non-monotonic dome structure around the critical pressure where the CDW is suppressed, strongly suggests an intimate interplay between the two states \cite{Chen_PRL,WuNature,FengNPJ,FengNatCom}. This complexity is mirrored in the intense debate surrounding the superconducting gap symmetry. The nuclear spin-lattice relaxation rate $1/T_1$ divided by $T$, $1/T_1T$, measured by nuclear quadrupole resonance (NQR) at ambient pressure shows a small Hebel-Slichter (coherence) peak just below $T_{\rm c}$ \cite{ZLi,WuNature,FengNatCom,Kitagawa}, which was taken as evidence for conventional $s$-wave superconductivity \cite{ZLi,Kitagawa}.  The robustness of $T_{\rm c}$ against nonmagnetic impurities is also indicative of $s$-wave superconductivity \cite{Roppongi}. The temperature dependence of the superfluid density and magnetic penetration depth below $T_{\rm c}$, measured by muon spin rotation ($\mu$SR) \cite{ZShan_muSR,GuptamuSR} and tunnel diode oscillator techniques \cite{Yuan, Roppongi}, has been explained by a two-gap $s$-wave model.

 In contrast, a growing body of evidence points toward an unconventional nature. At ambient pressure, this includes an in-plane two-fold rotational symmetry breaking observed in transport measurements \cite{Ni_CPL,XiangNatCom}, and a nodal gap component suggested by thermal conductivity \cite{HossainNatPhys} and NQR $1/T_1$ measurements \cite{FengNatCom} at low temperatures. Other exotic features such as a pair density wave (PDW) \cite{Chen_PDW,DengNature}, time-reversal symmetry breaking \cite{DengNature,LeNature}, and Majorana zero modes in vortex cores \cite{LiangPRX} have also been reported. In the high-pressure regime, $\mu$SR \cite{Guguchia_muSR} and NQR \cite{FengNatCom} experiments suggest a full gap, but an internal field was found by $\mu$SR, suggesting a time reversal symmetry broken superconducting state \cite{Guguchia_muSR}. The striking dichotomy regarding the gap symmetry at ambient pressure and the emergent phenomenon under high hydrostatic pressure underscore the complex nature of the pairing state in CsV$_3$Sb$_5$, highlighting the need for alternative experimental approaches.

In this Letter, we utilize the uniaxial pressure to tune the superconducting state. We measure  ac-susceptibility  (ac-$\chi$) and $^{121}$Sb-NQR to investigate the evolution of the gap in CsV$_3$Sb$_5$. We find that $T_{\rm c}$ is significantly enhanced by tensile strain, while $T_{\rm CDW}$ is insensitive to strain. At a maximum tensile strain of $\varepsilon$ = +0.90\%, through 1/$T_1T$ measurements, we discover remarkable double superconducting transitions. Below the upper superconducting onset temperature at $T_{\rm c1}$ = 3.6 K, 1/$T_1T$ decreases without a coherence peak and follows a power-law-like $T$ dependence, which is characteristic of a nodal state. In sharp contrast, a distinct coherence peak appears below the lower transition at $T_{\rm c2}$ = 3.0 K, indicating a conventional, nodeless $s$-wave gap. These results provide direct evidence for degenerate states with different gap symmetry that split under strain. Our work demonstrates a high tunability of superconductivity by uniaxial pressure.

\begin{figure}
\begin{center}
\includegraphics[width=0.9\linewidth]{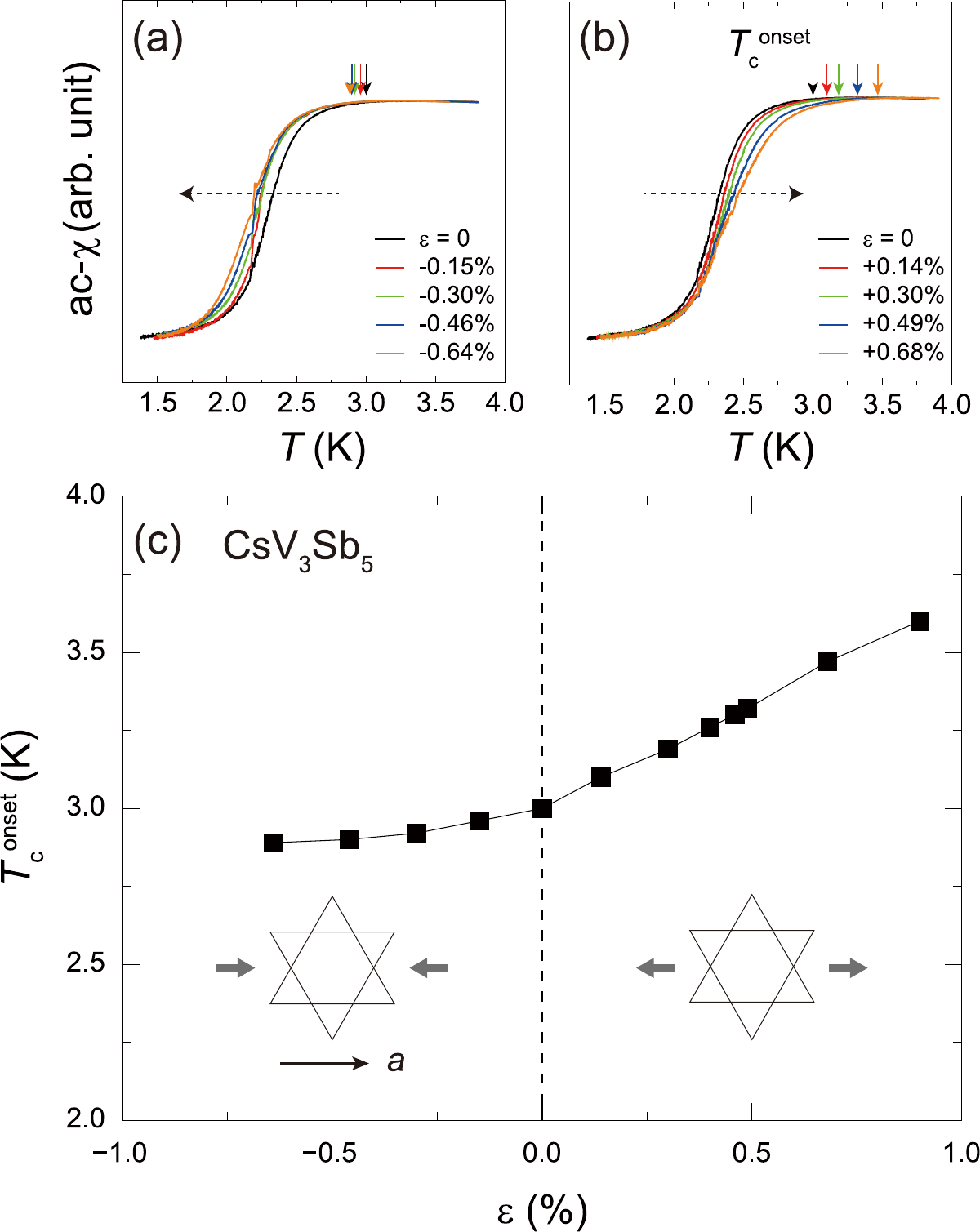}
\end{center}
    \caption{ Temperature dependence of the ac-$\chi$ under (a) compressive and (b) tensile strains, measured in zero magnetic field. Arrows indicate the superconducting transition onset, $T_{\rm c}^{\rm onset}(\varepsilon)$. (c) Strain dependence of  $T_{\rm c}^{\rm onset}(\varepsilon)$. The inset shows a schematic of the V kagome net, with arrows indicating the $a$-axis and the direction of the applied uniaxial pressure. The vertical dashed line marks the zero-strain position. }   
\end{figure}

High-quality single crystals of CsV$_3$Sb$_5$ were synthesized by the self-flux method \cite{Ortiz}. A thin, plate-like single crystal (1.00 $\times$ 0.60 $\times$ 0.08 mm$^3$) was strained along the crystallographic $a$-axis, determined by Laue diffraction. We used a homemade piezoelectric-driven strain cell [see Fig. 6 in the End Matter]. The strain ($\varepsilon$) is defined as $\varepsilon$(\%) = 100 $\times$ $\Delta$$L$/$L_0$, where $L_0$ is the initial length of the strained section \cite{TsukudaNatCom}. We performed $^{121}$Sb ($I$ = 5/2) NQR measurements on both the Sb(1) and Sb(2) sites in the CDW state at zero magnetic field \cite{LuoNPJ,FengNPJ}. $T_{\rm c}($$\varepsilon$$)$ was determined from ac-$\chi$, which was monitored via the resonance frequency of the NQR tank circuit. The $^{121}$Sb-NQR spectrum was taken by sweeping the rf frequency using a phase-coherent spectrometer. To obtain $T_1$, the recovery of the nuclear magnetization $M(t)$  after a saturation pulse was fitted by the theoretical function for $I$ = 5/2 and an asymmetry parameter $\eta$ = 0 \cite{recovery}, consistent with previous reports \cite{FengNPJ,FengNatCom,LuoNPJ}.

\begin{figure}
\begin{center}
\includegraphics[width=0.98\linewidth]{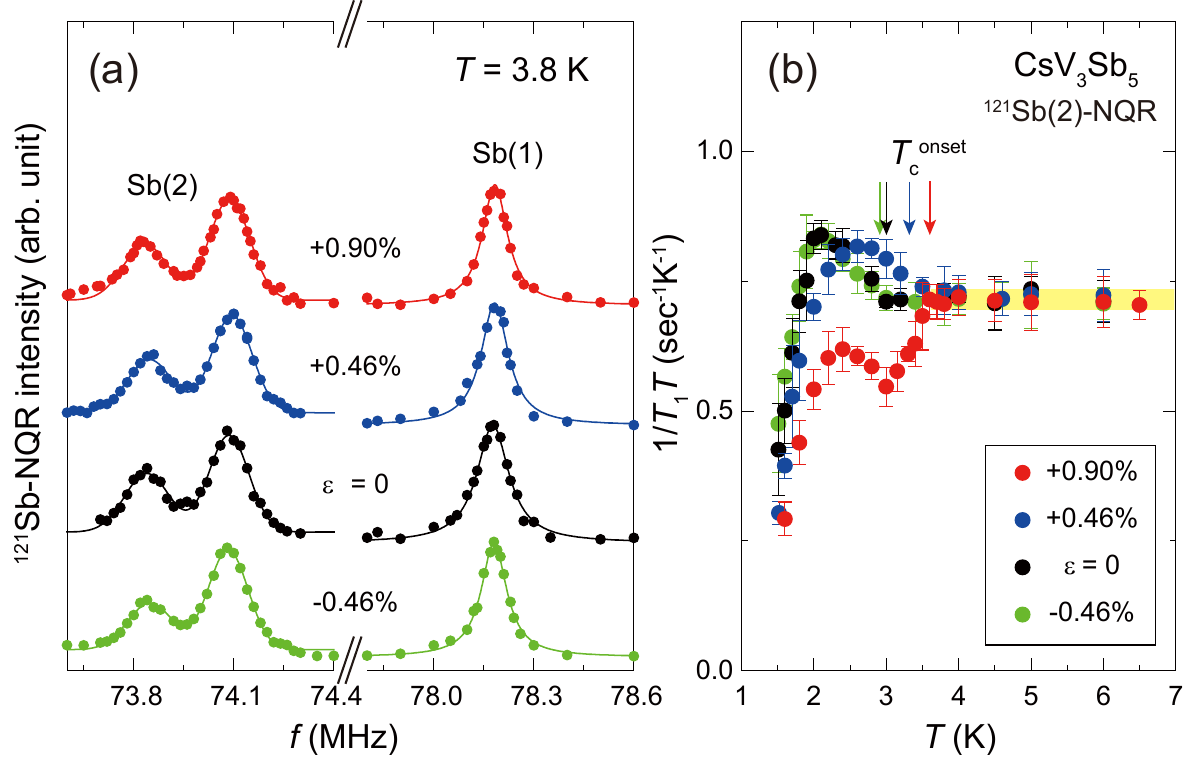}
\end{center}
    \caption{(a) Strain dependence of the $^{121}$Sb-NQR spectra in the CDW state at $T$ = 3.8 K. The spectra consist of the Sb(2) lines (left, $\sim 74$~MHz) and the Sb(1) line (right, $\sim 78.2$~MHz). The solid curves represent the best fits using a sum of two Gaussian functions for the Sb(2) site and a single Lorentzian function for the Sb(1) site. (b) Temperature dependence of the $^{121}$Sb-NQR $1/T_1T$ for various strains, measured at the Sb(2) site ($f = 74.1$~MHz). The horizontal line is a guide to the eye. Error bars represent the standard deviations from the fitting parameters. }   
\end{figure}

\begin{figure*}
\begin{center}
\includegraphics[width=0.95\linewidth]{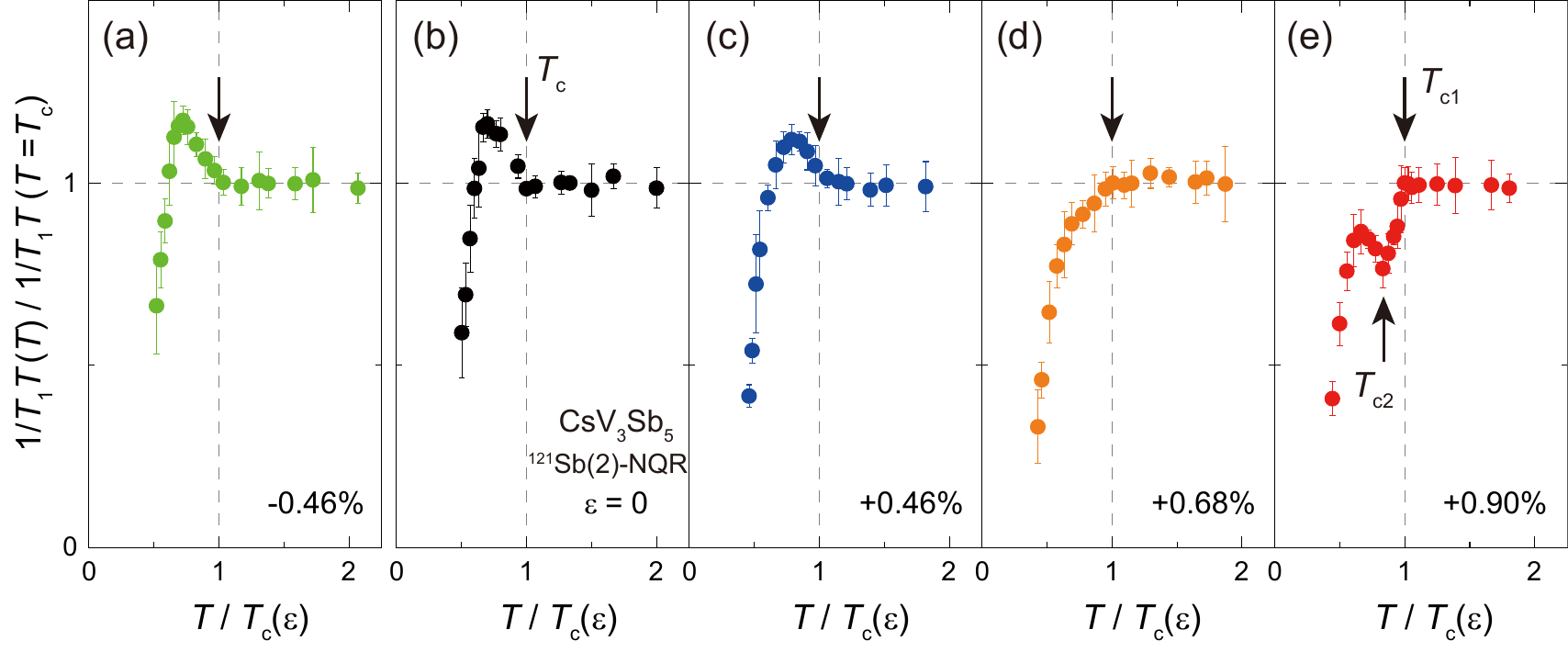}
\end{center}
    \caption{ Temperature dependence of the $^{121}$Sb(2)-NQR $1/T_1T$ under uniaxial strain. Data are shown for (a) $\varepsilon$  = -0.46\%, (b) 0, (c) +0.46\%, (d) +0.68\%, and (e) +0.90\%. The  $1/T_1T$  and temperature ($T$) are normalized by their respective values at $T_{\rm c}$$(\varepsilon)$. The dashed lines mark the normalized transition point at $T$/$T_{\rm c}$$(\varepsilon)$ = 1, and the solid arrows indicate the onset of the superconducting transition. Error bars represent the standard deviations from the fitting parameters.}   
\end{figure*}

Figures 1(a) and 1(b) show the temperature dependence of the ac-$\chi$ at various strains, while Fig. 1(c) plots the resulting superconducting transition onset temperature, $T_{\rm c}^{\rm onset}$, which is 3.0 K at zero strain ($\varepsilon = 0$), as a function of $\varepsilon$. While $T_{\rm c}^{\rm onset}$ slightly decreases under compression, it increases significantly under tension, reaching $T_{\rm c}^{\rm onset}$ = 3.6 K at $\varepsilon$ = +0.90\%. This continuous increase is qualitatively consistent with prior electrical resistivity measurements  \cite{Qian_PRB,Yang_CPB} (see Sec.1 in the Supplemental Material \cite{SM}) and contrasts sharply with materials like cuprates, where $d$-wave superconductivity is suppressed symmetrically by strain \cite{TsukudaNatCom}. This asymmetry may reflect that compression and tension are not crystallographically equivalent for the kagome lattice. Notably, the increase in $T_{\rm c}^{\rm onset}$ is accompanied by a substantial broadening of the superconducting transition width--an intriguing feature not captured by previous resistivity studies \cite{Qian_PRB,Yang_CPB}.

To microscopically investigate the electronic state, we performed $^{121}$Sb-NQR measurements [Figs.~2(a) and 2(b)]. As shown in Fig.~2(a), the NQR spectra at $T = 3.8$~K consist of two $^{121}$Sb(2) peaks around 73.8~MHz (low-frequency peak) and 74.1~MHz (high-frequency peak), as well as a distinct $^{121}$Sb(1) peak around 78.2~MHz. Consistent with previous NQR studies \cite{FengNPJ,LuoNPJ,FengNatCom}, the Sb(1) spectrum is substantially sharper than the Sb(2) spectra. More importantly, as explicitly demonstrated in Sec.~2 of the Supplemental Material \cite{SM}, the Sb(1) site is exceptionally sensitive to the local CDW order compared to the Sb(2) sites. Remarkably, the NQR frequencies for both the Sb(1) and Sb(2) sites show virtually no strain dependence (see Fig.~7 in the End Matter), and the temperature evolution of the Sb(1)-NQR frequency at $\varepsilon$ = +0.90\% perfectly tracks the zero-strain behavior (see Sec. 2 in the Supplemental Material \cite{SM}). Furthermore, as seen in Fig. 2(b), while the behavior below $T_{\rm c}^{\rm onset}$ changes markedly (as discussed later), $1/T_1T$ in the normal state (i.e., the CDW state) is completely unaffected by strain.

While these observations might raise the question of whether the macroscopic strain is effectively transferred to the sample, the linewidth of the highly sensitive Sb(1) peak provides direct microscopic proof to the contrary. Because the CDW transition breaks the sixfold rotational symmetry of the kagome lattice and forms three symmetry-equivalent twinned domains under zero strain \cite{Stahl_PRB2022}, the presence of these twin boundaries introduces a spatial distribution of the local electric field gradient (EFG), acting as a primary source of the NQR frequency inhomogeneity. Under applied strain, the linewidth of the Sb(1) peak exhibits a clear reduction (see Fig. 7 in the End Matter and Sec. 3 in the Supplemental Material \cite{SM}). This strain-induced spectral sharpening can be naturally understood as due to the detwinning of these domains, where the macroscopic alignment effectively homogenizes the local EFG and reduces the NQR linewidth. Therefore, the applied stress is genuinely transmitted to the bulk, confirming that the observed strain-independence of the NQR parameters reflects the highly robust, intrinsic nature of the CDW state. The successful transferring of  the strain to the sample is more clearly and directly shown by the drastic change of 1/$T_1$ in the superconducting state, as described later.

These findings are in contrast to previous reports; electrical resistivity measurements suggested a slight decrease in $T_{\rm CDW}$ \cite{Qian_PRB}, and angle-resolved photoemission spectroscopy (ARPES) experiments deduced a reduction of the CDW gap under tensile strain \cite{ChangNatCom}. The discrepancy may be understood by considering the specific nature of the probes. The observed shift of $T_{\rm CDW}$ in macroscopic transport might be associated with altered scattering dynamics rather than a thermodynamic suppression of the CDW phase. Specifically, the strain-induced detwinning suppresses domain-wall formation, which can naturally modify the resistivity anomaly used to define the transition. Consistent with this scenario, the reported resistivity peak becomes significantly broadened under strain \cite{Qian_PRB}. While momentum-resolved ARPES may detect anisotropic modifications of the CDW gap at specific $k$-points due to the broken rotational symmetry under uniaxial strain, our local NQR probes the overall real-space charge modulation amplitude.

1/$T_1T$ at the Sb site reflects the density of states (DOS) at the Fermi level \cite{FengNPJ}, while the NQR resonance frequency reflects the CDW order parameter \cite{FengNPJ,LuoNPJ}. The fact that both observables remain completely unchanged provides compelling evidence that the total DOS and the bulk CDW state are robust against strain. Our findings also demonstrate that the application of uniaxial pressure differs fundamentally from that of hydrostatic pressure, as it does not predominantly influence the superconducting state through the suppression of the CDW state \cite{Chen_PRL,WuNature,FengNPJ,FengNatCom}, which constitutes one of the core novelties of this work. Consequently, in the subsequent analysis, the impact of strain on the intrinsic CDW order parameter can be neglected.

Figure 3 shows the temperature dependence of 1/$T_1T$, normalized by its value at $T_{\rm c}(\varepsilon)$, plotted against the normalized temperature, $T/T_{\rm c}(\varepsilon)$. At $\varepsilon$ = 0 [Fig. 3(b)], a coherence peak is observed just below $T_{\rm c}$, consistent with previous reports \cite{ZLi,WuNature,FengNatCom}. This feature remains largely unaffected on the compressive side [Fig. 3(a)], consistent with the small change in  $T_{\rm c}(\varepsilon)$. In contrast, a clear evolution is observed under tension [Figs. 3(c)-3(e)]. At $\varepsilon$ = +0.46\%, the coherence peak is slightly suppressed, and by +0.68\%, it disappears completely. The complete suppression of the coherence peak signifies a substantial increase in the overall superconducting gap anisotropy, as gap anisotropy is known to suppress the coherence peak \cite{MukudaCeRu2}. This could arise either from a highly anisotropic, yet still nodeless, $s$-wave gap, or from the emergence of a nodal component. Meanwhile, the remaining convex curvature suggests that a nodeless gap component still governs the overall quasiparticle response. 

Upon applying further tension to $\varepsilon$ = +0.90\%, a striking change appears. As seen in Fig. 3(e), 1/$T_1T$ steeply decreases just below the upper transition at $T_{\rm c1}$ $\approx$ 3.6~K, which coincides with $T_{\rm c}^{\rm onset}$ determined from ac-$\chi$. This lack of a coherence peak is characteristic of nodal superconductivity \cite{AsayamaReview}. In sharp contrast, this is immediately followed by a distinct coherence peak just below a second, lower transition at $T_{\rm c2}$ $\approx$ 3.0 K. This result unambiguously demonstrates double superconducting transitions involving two distinct gap symmetries: an emergent nodal gap and the nodeless $s$-wave gap observed at zero strain.

A previous NQR study at ambient pressure down to very low temperature suggested a multi-gap state of nodeless ($s$)  and nodal ($d$) components with a spectral weight ratio of approximately 9 : 1, with gap magnitudes of $2\Delta_s(0)/k_{\rm B}T_{\rm c} = 4.0$ and $2\Delta_d(0)/k_{\rm B}T_{\rm c} = 1.26$, respectively  \cite{FengNatCom}. To capture the evolution of this superconducting state under strain, Fig. 4 represents the temperature dependence of $1/T_1$ at $\varepsilon$ = 0 and +0.90\%. We applied the same two-component ($s+d$ wave) model \cite{FengNatCom} to the data at  $\varepsilon$ = +0.90\% below $T_{\rm c}^{\rm onset}$. By assuming the $d$-wave and $s$-wave gaps open at $T_{\rm c1}$ = 3.6~K and $T_{\rm c2}$ = 3.0~K, respectively, the best-fit parameters are obtained as $2\Delta_{d}(0)/k_{\rm B}T_{\rm c1} = 7.95$ and $2\Delta_{s}(0)/k_{\rm B}T_{\rm c2} = 2.81$, with a $d$-wave volume fraction $w \approx 0.26$ (see the End Matter and Sec. 4 in the Supplemental Material \cite{SM} for details).  
 
\begin{figure}
 \begin{center}
\includegraphics[width=0.95\linewidth]{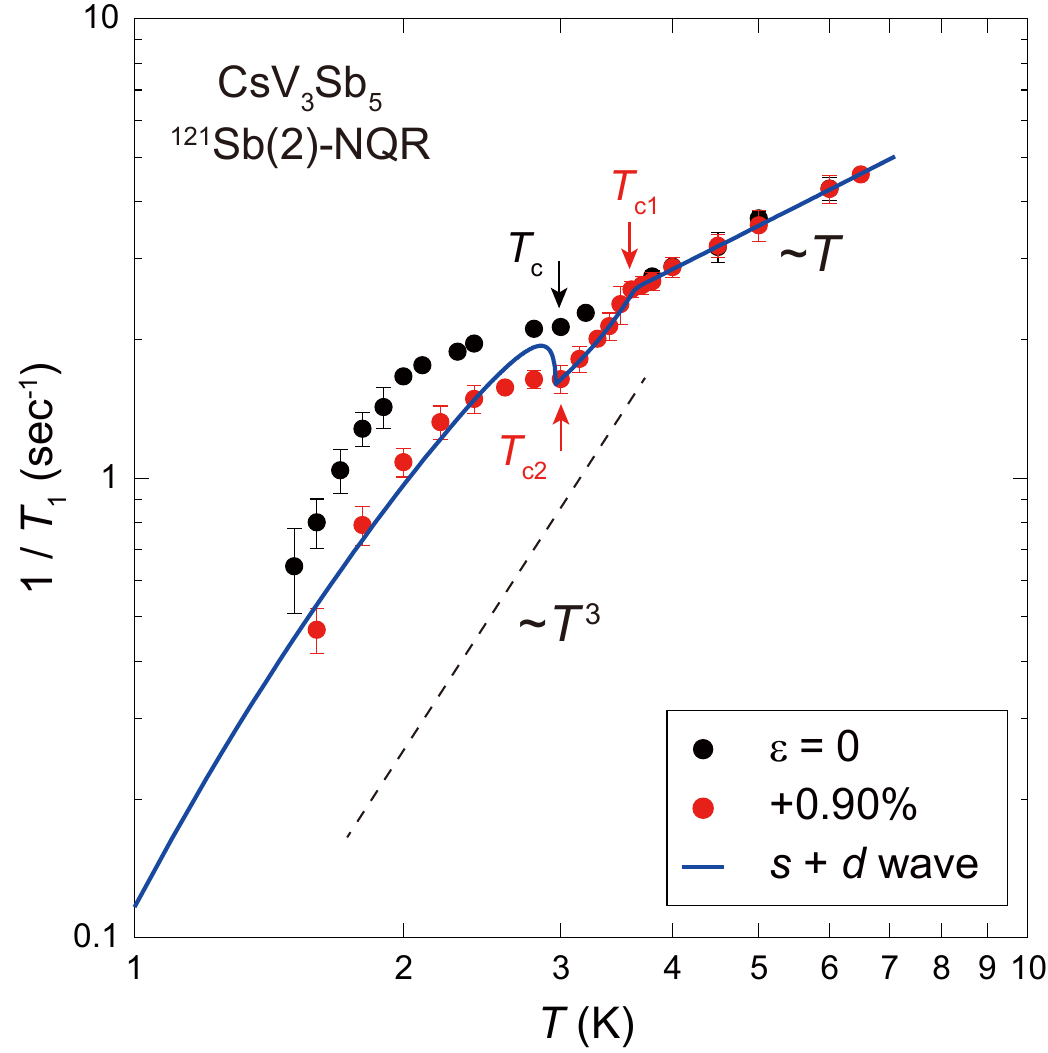}
\end{center}
    \caption{Temperature dependence of the $^{121}$Sb(2)-NQR $1/T_1$ at $\varepsilon$ = 0 and +0.90\%. Arrows indicate the respective superconducting transition temperatures, $T_{\rm c}$. The solid and dashed lines show linear ($\sim$$T$) and cubic ($\sim$$T^3$) dependencies above and below $T_{\rm c}$, respectively, as guides to the eye. The solid curve below $T_{\rm c1}$ represents the fitting result with a two-component ($s+d$ wave) model for $\varepsilon = +0.90\%$ (see text). Error bars represent the standard deviations from the fitting parameters.}  
\end{figure}

This quantitative analysis necessitates a refinement of the strain effect picture. The primary effect of strain is identified as a selective and drastic enhancement of the nodal pairing channel, rather than a mere tuning of individual transition temperatures. Specifically, tensile strain boosts the $d$-wave gap ratio $2\Delta_d(0)/k_{\rm B}T_{\rm c1}$ from 1.26 at zero strain to 7.95 at $\varepsilon$ = +0.90\%, and expands its volume fraction $w$ from $\sim$10 \% to 26 \%. Since the DOS remains unchanged, this result indicates a direct strengthening of the pairing interaction in the $d$-wave channel.

We therefore conclude that CsV$_3$Sb$_5$ is a multi-gap superconductor with both nodeless ($s$-wave) and nodal ($d$-wave) components. The relatively small fraction of the nodal component at ambient pressure ($\sim$ 10\%) likely explains the historical difficulty in resolving the gap symmetry, which has been the source of the longstanding debate.

Our results place strong constraints on possible unconventional pairing scenarios at ambient pressure. A chiral superconducting state, which breaks time-reversal symmetry (e.g.,  $d_{x^2-y^2}+id_{xy}$), should show a $T_{\rm c}$ enhancement under both compression and tension as strain lifts the degeneracy \cite{SigristUeda}, a scenario that contradicts our findings. Another proposed state is a pair density wave (PDW) \cite{Chen_PDW,DengNature}, a state where the superconducting order parameter is spatially modulated \cite{Fradkin}. It was theoretically proposed that a PDW state may give rise to a nodal-like gap on V-orbitals \cite{JXYin_PRB}, which would be consistent with the nodal gap we observe at $T_{\rm c1}$. However, a PDW state would also be expected to induce a spatial modulation of the local electronic environment. This should manifest as a discernible broadening or splitting of the NQR spectrum below $T_{\rm c}$.

As shown in Fig. 8 in the End Matter, the NQR spectra for both the Sb(1) and Sb(2) sites exhibit no additional broadening or splitting upon cooling well below $T_{\rm c1}$ at a tensile strain of $\varepsilon = +0.90$\%. The absence of static EFG modulation, even at the highly sensitive Sb(1) site, does not lend support for an emergence of a static PDW order coexisting with superconductivity in the strained state.

\begin{figure}
\begin{center}
\includegraphics[width=0.95\linewidth]{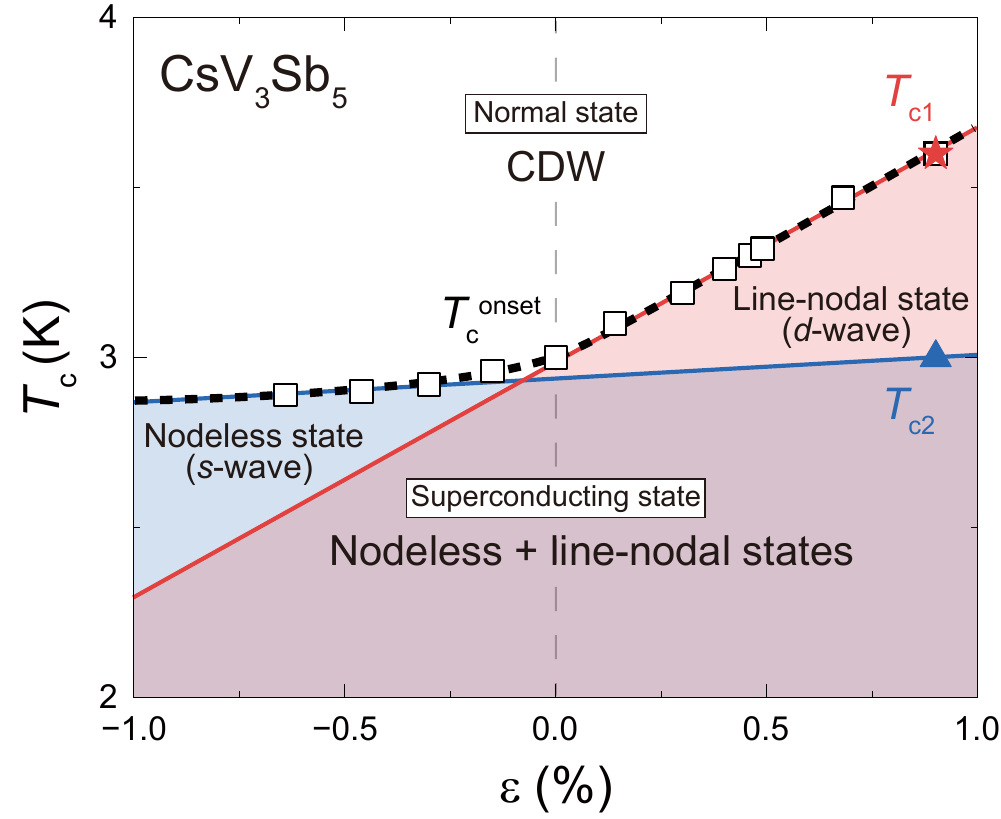}
\end{center}
    \caption{Strain-temperature phase diagram of superconductivity in CsV$_3$Sb$_5$ at zero magnetic field. Open squares show $T_{\rm c}^{\rm onset}$ determined from ac-$\chi$. The solid star and triangles represent the upper ($T_{\rm c1}$) and lower ($T_{\rm c2}$) transition temperatures, respectively, as determined from the temperature dependence of $1/T_1T$. The dotted line is a guide to the eye following $T_{\rm c}^{\rm onset}$. Solid lines trace the boundaries for the emergent line-nodal state (possibly $d$-wave) and the nodeless state ($s$-wave) (see text). Vertical dashed line indicates $\varepsilon$ = 0.}   
\end{figure}

Finally, we show the strain-temperature phase diagram in Fig. 5 as deduced from our measurements. This phase diagram provides a qualitative explanation for the broadening of the superconducting transition observed in the ac-$\chi$ under tension; it can be understood as originating from the splitting of $T_{\rm c}$ into two distinct transitions. Furthermore, even at ambient pressure, because the $s$-wave and $d$-wave states are nearly (but not completely) degenerate, the finite transition width intrinsically reflects the overlapping contributions from these two states, which explicitly split upon applying strain.  Our $1/T_1T$ results directly confirm this picture. At the upper transition at $T_{\rm c1}$, 1/$T_1$ lacks a coherence peak and exhibits a power-law-like $T$ dependence (Fig. 4), both of which are characteristic of a line-nodal state, while at the lower transition at $T_{\rm c2}$, 1/$T_1$ shows a distinct coherence peak, confirming the nodeless state.

In this degenerate-states scenario, the two states have distinct responses to strain: the nodeless ($s$-wave) state is insensitive to strain, a behavior consistent with its isotropic nature. In contrast, the line-nodal ($d$-wave) state exhibits a strong sensitivity to strain, a characteristic previously demonstrated in cuprates \cite{TsukudaNatCom}. An ARPES study has suggested that the vHs moves closer to the Fermi level (from -80 meV to -40 meV) when going from compression to tension, +1\% \cite{ChangNatCom}. Indeed, the position of the vHs relative to the Fermi level exhibits an asymmetric response to strain that tracks the behavior of $T_{\rm c1}$ \cite{ChangNatCom} and can have a correlation with the line-nodal superconductivity we observed.

 However, the vHs does not actually reach the Fermi level. Our normal-state 1/$T_1T$ results confirm that the enhancement of $T_{\rm c}$ is not accompanied by an increase in the total DOS or spin fluctuations.  This explicitly demonstrates that the strain-induced enhancement of the line-nodal superconductivity is driven by a direct strengthening of the anisotropic pairing interaction, rather than a simple increase in the DOS [$N(E_{\rm F})$]. This observation perfectly aligns with the theoretical prediction that lattice deformation directly modifies the electronic structure near the saddle points ($M$ points) at the Fermi level \cite{Fernandes_PRB}. Crucially, our finding that the $T_{\rm c}$ enhancement is completely independent of the in-plane strain direction (see Sec. 5 in the Supplemental Material \cite{SM}) strongly supports this picture; it implies that the local modulation of the V-V bonds, rather than macroscopic in-plane anisotropy, primarily tunes these $M$-point states. While Ritz {\it et al.} \cite{Fernandes_PRB} discussed this in the context of electron-phonon coupling, the sensitivity of the $M$-point states to such local strain implies that the associated scattering channels are highly tunable. Indeed, our observation of the significant increase in the $d$-wave fraction under strain suggests that the strain-induced modification of the Fermi surface enhances the inter-valley scattering responsible for the anisotropic pairing. This conclusion is further supported by the ARPES study itself, which observed significant electron-phonon coupling kinks, highlighting the indispensable role of lattice degrees of freedom \cite{ChangNatCom}. Further studies are needed to elucidate the microscopic origin of this pairing enhancement.

In conclusion, we have investigated the superconducting state in CsV$_3$Sb$_5$ under in-situ uniaxial pressure. Our key finding is that superconductivity can be enhanced independently of the robust CDW order. Under sufficient tensile strain ($\varepsilon$ = +0.90\%), we discovered remarkable double superconducting transitions, providing unambiguous evidence for two distinct pairing states: a selectively enhanced line-nodal state at a higher $T_{\rm c1}$ = 3.6~K and a nodeless $s$-wave state at a lower $T_{\rm c2}$ = 3.0~K. Our results demonstrate that CsV$_3$Sb$_5$ hosts multiple competing superconducting pairing channels, and that strain acts as a tuning parameter to stabilize the nodal component,  providing a distinct pathway to understanding the intrinsic pairing mechanism in this fascinating kagome superconductor.

\begin{acknowledgments}
$Acknowledgments$ --This work was supported by JSPS KAKENHI Grant Numbers JP19H00657, JP19K03747, JP23K03323, JP26K00657, and JP26K07015; the JSPS Program for Forming Japan's Peak Research Universities (J-PEAKS) Grant Number JPJS00420230010; research grants from the Murata Science and Education Foundation, the Electric Technology Research Foundation of Chugoku, and the Okayama Foundation for Science and Technology; and the CAS PIFI program (2024PG0003).
\end{acknowledgments}



\section{End Matter}

$Strain$ $cell$ $and$ $method$ --Figure 6(a) shows a photograph of the homemade piezoelectric-driven strain cell used in this study \cite{TsukudaNatCom}. Following the design pioneered by Hicks $et$ $al$. \cite{Hicks} the cell is assembled from three pure titanium (99.5\%) components, fabricated by electric discharge machining, and three commercial piezoelectric actuators (PI, P-885.51) controlled by a Razorbill Instruments RP100 power supply. Two actuators are embedded in the outer components and one in the central component. A homemade parallel-plate capacitor (gap $<$ 0.1~mm, $C$ $\approx$ 1.5 pF at room temperature) is fixed to the cell, and its capacitance is monitored by a Keysight E4980AL LCR meter to determine the displacement. A single crystal plate, oriented with its $a$-axis along the strain direction, is mounted across a 1.00 mm gap and secured at both ends with small droplets of degassed epoxy (STYCAST 2850FTJ). The NQR coil (30 turns of 0.03 mm Cu wire) is hand-wound on a paraffin paper former, placed directly over the crystal, and fixed in place with GE7031 varnish. To calibrate the cell and confirm the crystal's elastic response, we measured the strain ($\varepsilon$) as a function of the voltage applied to the actuators at $T$ = 2~K [Fig.~6(b)]. The observed linear relationship confirms that the crystal deformation is elastic and obeys Hooke's law, yielding the relation $\varepsilon(\%) \approx 0.00461 \cdot V_{\rm piezo}$.  Crucially, we explicitly confirmed this perfectly elastic and reversible relation {\it in situ} prior to every $T_{\rm c}$ and NQR measurement at each temperature, guaranteeing that no irreversible deformation occurred during the experiments. This $in$ $situ$ capacitance monitoring provides a direct and quantitative measure of the actual macroscopic deformation, ensuring that the large strain of up to +0.90\% is faithfully transferred to the bulk crystal.

\begin{figure}[h]
	\begin{center}
		\includegraphics[width=0.8\linewidth]{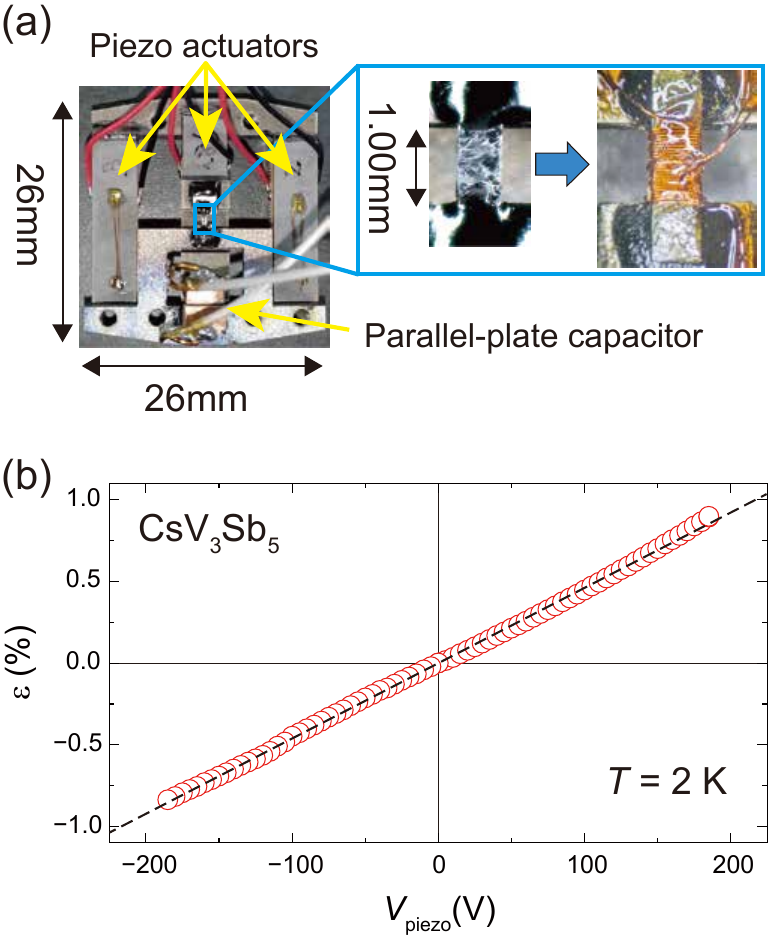}
	\end{center}
	\caption{Experimental setup for the in-situ uniaxial strain NQR measurements. (a) Photograph of the homemade piezoelectric-driven strain cell, showing the mounted single crystal and the NQR coil. (b) Strain ($\varepsilon$) as a function of the voltage applied to the piezoelectric actuators ($V_{\rm piezo}$), measured at $T$ = 2~K. The dashed line is a linear fit to the data.}   
\end{figure}

\begin{figure}
	\begin{center}
		\includegraphics[width=1\linewidth]{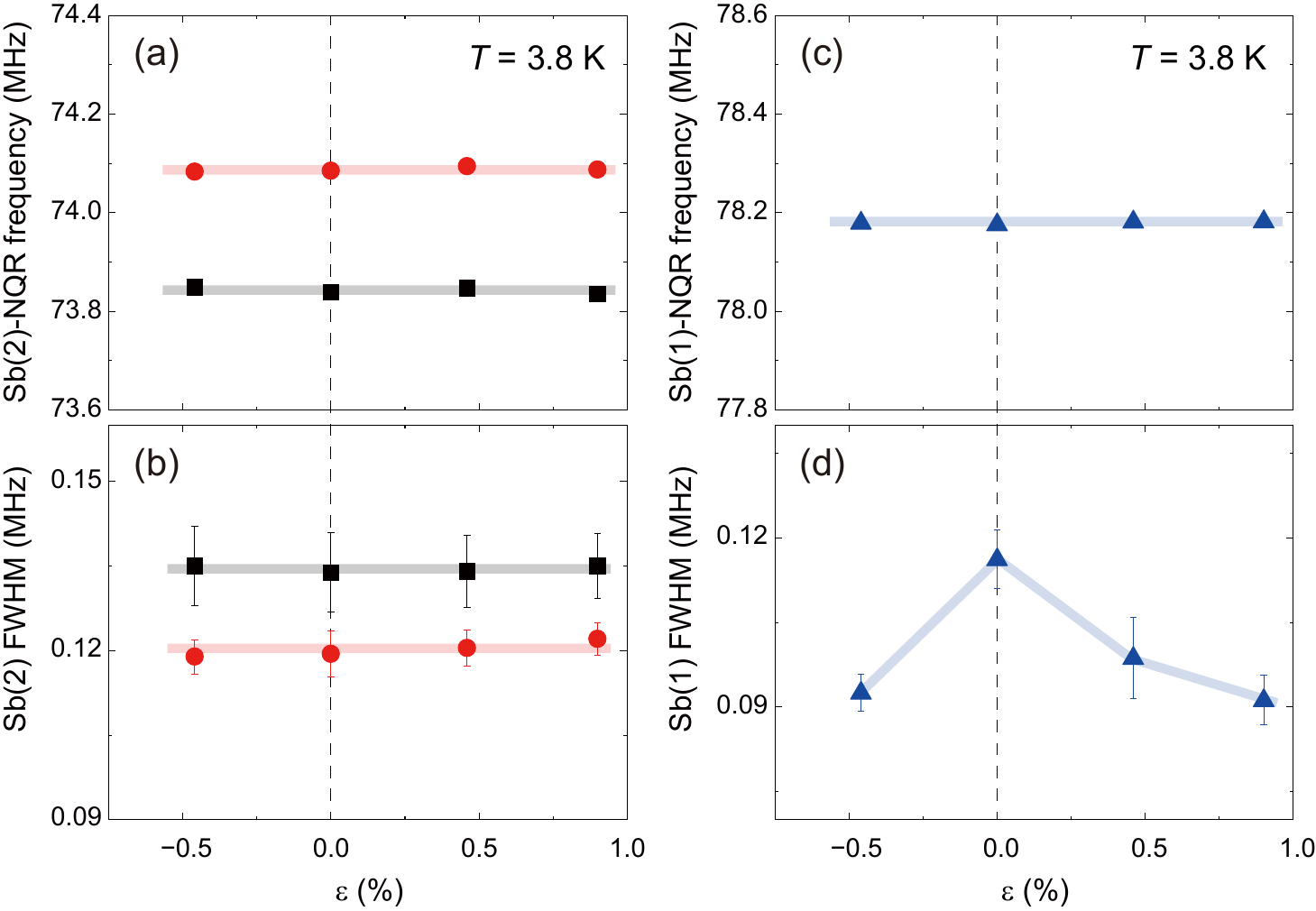}
	\end{center}
	\caption{Strain dependence of the NQR peak frequencies and the full-widths at half maximum (FWHM) at $T = 3.8$~K. (a), (b) Parameters for the Sb(2) site obtained by a two-Gaussian fit. Black solid squares and red solid circles represent the low-frequency and high-frequency peaks, respectively. (c), (d) Parameters for the Sb(1) site obtained by a single-Lorentzian fit. Blue solid triangles denote the single Sb(1) peak. Error bars represent the standard deviations from the fitting parameters. The vertical dashed lines indicate zero strain, and the thick shaded lines are guides to the eye.
	}  
\end{figure}

$Strain$ $dependence$ $of$ $the$ $NQR$ $spectra$ --Figure 7 shows the strain dependence of the peak frequencies and the full-widths at half maximum (FWHM) extracted from the NQR spectra in Fig.~2(a). These parameters were obtained using a two-Gaussian fit for the Sb(2) site [Figs.~7(a) and 7(b)] and a single-Lorentzian fit for the Sb(1) site [Figs.~7(c) and 7(d)].

Notably, while the resonance frequency of the Sb(1) site remains completely unchanged regardless of the applied strain [Fig.~7(c)], its FWHM exhibits a clear reduction under both tensile ($\varepsilon > 0$) and compressive ($\varepsilon < 0$) strains [Fig.~7(d)]. At zero strain, the presence of multiple twinned domains introduces a spatial distribution of the local electric field gradient (EFG), which broadens the NQR line. Therefore, this symmetric line narrowing provides compelling microscopic evidence for the detwinning of the CDW domains; the applied uniaxial strain lifts the orientational degeneracy, aligning the domains and thereby homogenizing the local EFG. It explicitly confirms that the applied uniaxial strain effectively and homogeneously penetrates the sample, while the intrinsic CDW order parameter itself is robust and unaffected. The fact that the Sb(1) line remains sharp even after the strain is released, indicative of a domain training effect, is further detailed in Sec. 3 in the Supplemental Material \cite{SM}.

$Analysis$ $of$ $the$ $temperature$ $dependence$ $of$ $1/T_1$ $below$ $T_{\rm c}$--The temperature dependence of $^{121}$Sb-NQR $1/T_1$ in Fig. 4 is re-plotted from the $1/T_1T$ data presented in Fig. 2(b). Following the analysis in Ref.~\cite{FengNatCom}, we employed a two-component ($s+d$) model to reproduce the temperature dependence of $1/T_1$ in the superconducting state. The normalized relaxation rate below $T_{\rm c}$ is expressed as $\frac{T_{1}(T_{\rm c})}{T_{1,{\rm SC}}} = \frac{2}{k_{\rm B}T_{\rm c}} \int_0^{\infty} \left[ N(E)^2 + M(E)^2 \right] f(E)[1-f(E)]dE$, where $N(E) = N(E_{\rm F}) E / \sqrt{E^2 - \Delta^2}$ is the superconducting DOS, and $M(E) = N(E_{\rm F}) \Delta / \sqrt{E^2 - \Delta^2}$ corresponds to the anomalous DOS arising from the coherence factor. Here, $N(E_{\rm F})$ is the DOS in the normal state, and $f(E)$ is the Fermi distribution function. To account for quasiparticle damping, we convolved $N(E)$ and $M(E)$ with a rectangular broadening function of width $2\delta$ and height $1/2\delta$~\cite{Hebel}. For the two-component model, we assume that the total effective DOS is a weighted sum of the two contributions: $N_{\rm tot}(E) = w N_d(E) + (1-w) N_s(E)$, where $N_d$ and $N_s$ denote the DOS for the $d$-wave and $s$-wave components, respectively. For the $d$-wave component, we used the line-nodal gap function $\Delta_d(\phi) = \Delta_0^d \cos(2\phi)$. For the $s$-wave component, we assumed a simple isotropic gap $\Delta_s(\phi) = \Delta_0^s$, neglecting in-plane anisotropy. The solid curve in Fig.~4 represents the calculation result using the best-fit parameters: $2\Delta_d(0)/k_{\rm B}T_{\rm c1} = 7.95$, $2\Delta_s(0)/k_{\rm B}T_{\rm c2} = 2.81$, a $d$-wave volume fraction of $w \approx 0.26$, and a broadening parameter $b \equiv \delta/\Delta_0^s \approx 0.56$ for the $s$-wave component.  

\begin{figure}
	\begin{center}
		\includegraphics[width=1\linewidth]{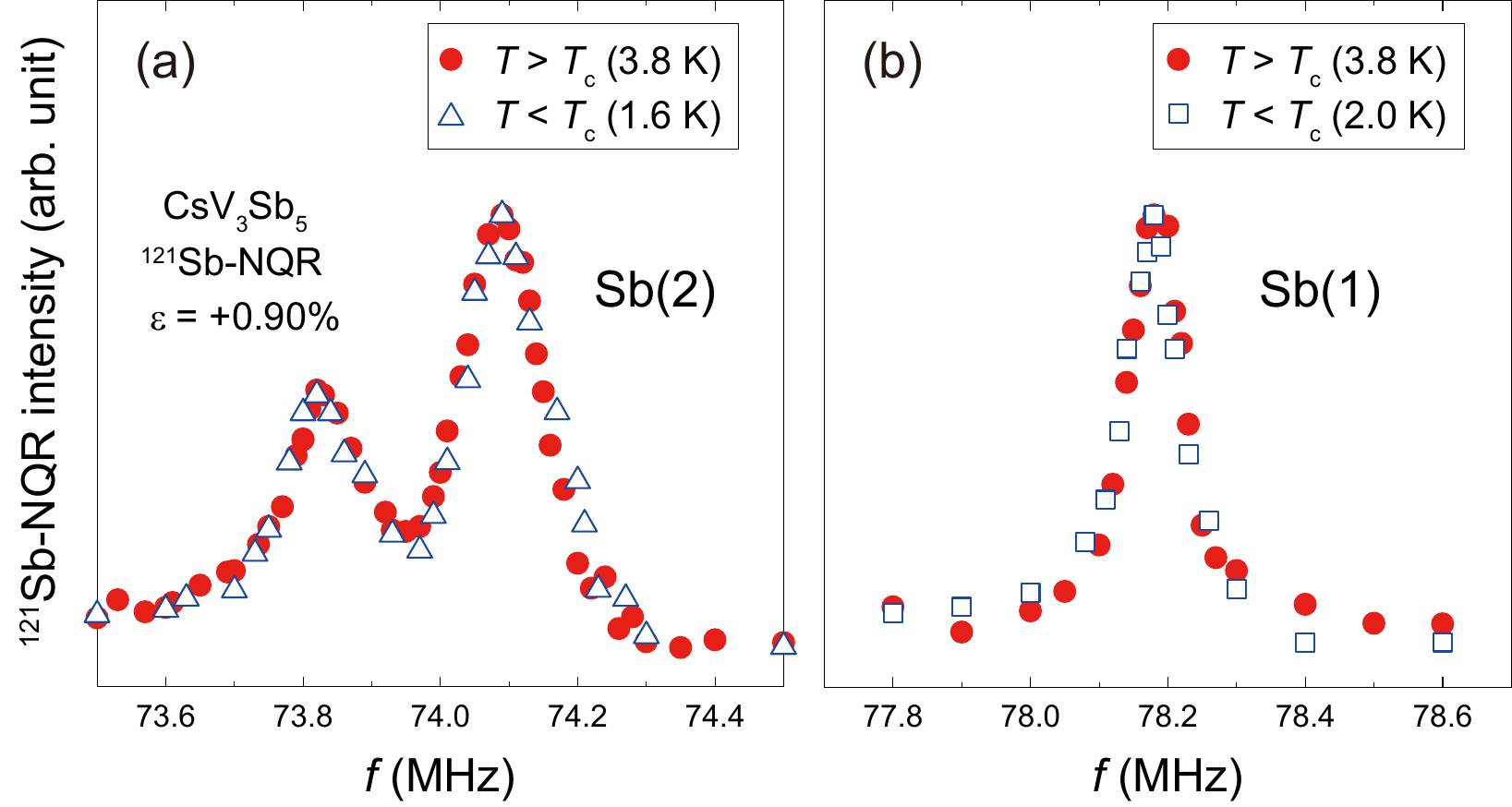}
	\end{center}
	\caption{The $^{121}$Sb-NQR spectra at a tensile strain of  $\varepsilon$ = +0.90\%  for (a) the Sb(2) site and (b) the Sb(1) site. Measurements were performed in the normal state ($T = 3.8$~K, above $T_{\rm c1}$; solid red circles) and in the superconducting state well below $T_{\rm c1}$, specifically at $T = 1.6$~K (open blue triangles) for Sb(2) and at $T = 2.0$~K (open blue squares) for Sb(1).}  
\end{figure}

{\it The NQR spectrum in the coexistent CDW and superconducting state}---Figure 8 shows the $^{121}$Sb-NQR spectra for both the Sb(2) [Fig.~8(a)] and Sb(1) [Fig.~8(b)] sites at a tensile strain of $\varepsilon = +0.90$\%. The measurements were performed at two temperatures: in the normal state ($T = 3.8$~K, above $T_{\rm c1} = 3.6$~K) and well below $T_{\rm c1}$ ($T = 1.6$~K for Sb(2) and $T = 2.0$~K for Sb(1)).

As is evident from the figure, neither spectrum shows discernible changes upon entering the superconducting state. The absence of any additional line broadening or splitting, even at the highly sensitive Sb(1) site, tightly constrains any static EFG modulation. This observation does not lend support for the emergence of a static PDW order coexisting with superconductivity in the strained state.


\begin{thebibliography}{9}

  \bibitem{Dagotto}
    E. Dagotto, Science {\bf 309}, 257 (2005).
  
   \bibitem{Keimer}
       B. Keimer $et$ $al$., Nature {\bf 518}, 179 (2015). 
  
  \bibitem{FernandesFe}
  R. M. Fernandes $et$ $al$., Nature {\bf 601}, 35 (2022). 
  
  \bibitem{KitahashiNatCom}
      S. Kawasaki $et$ $al$., Nat. Commun. {\bf 8}, 1267 (2017).
      
     \bibitem{ZhouNatCom}
         R. Zhou $et$ $al$., Nat Commun {\bf 4}, 2265 (2013).
 
   
   \bibitem{FengSL}
   Z. Feng $et$ $al$., Chin. Phys. Lett. {\bf 34}, 077502 (2017).
   
\bibitem{Lee_NatPhys}
A. T. Breidenbach $et$ $al$., Nat. Phys. {\bf 21}, 1957 (2025).

  
  \bibitem{QSL}
  C. Broholm $et$ $al$., Science {\bf 367},  eaay0668 (2020). 

   
  
\bibitem{Ortiz}
B. R. Ortiz $et$ $al$., Phys. Rev. Lett. {\bf 125}, 247002 (2020).

\bibitem{OrtizReview}
S. D. Wilson and B. R. Ortiz, Nat. Rev. Mater. {\bf 9}, 420 (2024).


\bibitem{ZWangSTM}
Z. Wang $et$ $al$., Phys. Rev. B {\bf 104}, 075148  (2021).
 
\bibitem{OrtizPRX}
B. R. Ortiz $et$ $al$., Phys. Rev. X {\bf 11}, 041030 (2021).

\bibitem{ZShan_muSR}
Z. Shan $et$ $al$., Phys. Rev. Research {\bf 4}, 033145 (2022).

\bibitem{LuoNPJ}
J. Luo $et$ $al$., npj Quantum Mater. {\bf 7}, 30 (2022). 

\bibitem{TazaiPNAS}
R. Tazai $et$ $al$., PNAS {\bf 121}, e2303476121 (2024).



\bibitem{Chen_PRL}
K. Y. Chen $et$ $al$., Phys. Rev. Lett. {\bf 126}, 247001 (2021).




\bibitem{FengNPJ}
X. Y. Feng $et$ $al$., npj Quantum Mater. {\bf 8}, 23 (2023). 

\bibitem{WuNature}
L. Zheng $et$ $al$., Nature {\bf 611}, 682 (2022). 

\bibitem{FengNatCom}
X. Y. Feng $et$ $al$., Nat. Commun. {\bf 16}, 3643 (2025). 

\bibitem{ZLi}
C. Mu $et$ $al$., Chin. Phys. Lett. {\bf 38}, 077402 (2021). 


\bibitem{Kitagawa}
M. Shibata $et$ $al$., Commun. Phys. {\bf 8}, 298 (2025). 

\bibitem{Roppongi}
M. Roppongi $et$ $al$., Nat. Commun. {\bf 14}, 667 (2023). 


\bibitem{GuptamuSR}
R. Gupta $et$ $al$., npj Quantum Mater. {\bf 7}, 49 (2022). 

\bibitem{Yuan}
W. Duan $et$ $al$., Sci. China Phys. Mech. Astron. {\bf 64}, 107462 (2021).



\bibitem{Ni_CPL}
S. Ni $et$ $al$., Chin. Phys. Lett. {\bf 38}, 057403 (2021).

\bibitem{XiangNatCom}
Y. Xiang $et$ $al$., Nat. Commun. {\bf 12}, 6727 (2021). 

\bibitem{HossainNatPhys}
Md S. Hossain $et$ $al$., Nat. Phys. {\bf 21}, 556 (2025). 


\bibitem{Chen_PDW}
H. Chen $et$ $al$., Nature {\bf 599}, 222 (2021). 




\bibitem{DengNature}
H. Deng $et$ $al$., Nature {\bf 632}, 775 (2024). 

\bibitem{LeNature}
T. Le $et$ $al$., Nature {\bf 630}, 64 (2024). 


\bibitem{LiangPRX}
Z. Liang $et$ $al$., Phys. Rev. X {\bf 11}, 031026 (2021).


\bibitem{Guguchia_muSR}
Z. Guguchia $et$ $al$., Nat. Commun. {\bf 14}, 153 (2023). 

\bibitem{TsukudaNatCom}
S. Kawasaki $et$ $al$., Nat. Commun. {\bf 15}, 5082 (2024). 

\bibitem{recovery}
D. E. MacLaughlin $et$ $al$., Phys. Rev. B {\bf 4}, 60 (1971).

\bibitem{Qian_PRB}
T. Qian $et$ $al$., Phys. Rev B {\bf 104}, 144506 (2021).

\bibitem{Yang_CPB}
X. Yang $et$ $al$., Chin. Phys. B {\bf 32}, 127101 (2023).

\bibitem{SM}
See Supplemental Material at https://journals.aps.org/authors/supplemental-material-instructions 

\bibitem{Stahl_PRB2022}
Q. Stahl $et$ $al$., Phys. Rev. B {\bf 105}, 195136 (2022).

\bibitem{ChangNatCom}
C. Lin $et$ $al$., Nat. Commun. {\bf 15}, 10466 (2024). 

\bibitem{MukudaCeRu2}
H. Mukuda $et$ $al$., J. Phys. Soc. Jpn. {\bf 67}, 2101 (1998).

\bibitem{AsayamaReview}
K. Asayama $et$ $al$., Prog. Nucl. Mag. Res. Sp. {\bf 28}, 221 (1996).

\bibitem{Fernandes_PRB}
E. T. Ritz $et$ $al$., Phys. Rev. B {\bf 108}, L100510 (2023). 

\bibitem{SigristUeda}
M. Sigrist and K. Ueda,  Rev. Mod. Phys. {\bf 63}, 239 (1991).

\bibitem{Fradkin}
 E. Fradkin, S. A. Kivelson, and J. M. Tranquada,  Rev. Mod. Phys. {\bf 87}, 457 (2015).

\bibitem{JXYin_PRB}
M. Yao $et$ $al$., Phys. Rev. B {\bf 111}, 094505 (2025).


\bibitem{Hicks}
C. W. Hicks $et$ $al$., Rev. Sci. Instrum. {\bf 85}, 065003 (2014).

\bibitem{Hebel}
 L. C. Hebel, Phys. Rev. {\bf 116}, 79 (1959).



\end{thebibliography}
\end{document}